\documentclass[a4paper]{article}

\usepackage{epsfig}
\usepackage{subcaption}
\usepackage{calc}
\usepackage{amssymb}
\usepackage{amstext}
\usepackage{amsmath}
\usepackage{amsthm}
\usepackage{multicol}
\usepackage{pslatex}
\usepackage{apalike}
\usepackage{algorithm2e}
\usepackage[bottom]{footmisc}
\usepackage{SCITEPRESS}     
\usepackage{multirow}

\begin{document}

\title{Optimizing Intensive Database Tasks Through Caching Proxy Mechanisms}

\author{\authorname{Ionuț-Alex Moise\sup{1} and Alexandra Băicoianu\sup{2}\orcidAuthor{0000-0002-1264-3404}}
\affiliation{\sup{1}Faculty of Mathematics and Computer Science, Transilvania University of Brașov, Brașov, Romania}
\affiliation{\sup{2}Faculty of Mathematics and Computer Science, Department of Mathematics and Computer Science, Transilvania University of Brașov, Brașov, Romania}
\email{alex.moise@student.unitbv.ro, a.baicoianu@unitbv.ro}
}

\abstract{Web caching is essential for the World Wide Web, saving processing power, bandwidth, and reducing latency. Many proxy caching solutions focus on buffering data from the main server, neglecting cacheable information meant for server writes. Existing systems addressing this issue are often intrusive, requiring modifications to the main application for integration. We identify opportunities for enhancement in conventional caching proxies. This paper explores, designs, and implements a potential prototype for such an application. Our focus is on harnessing a faster bulk-data-write approach compared to single-data-write within the context of relational databases. If a (upload) request matches a specified cacheable URL, then the data will be extracted and buffered on the local disk for later bulk-write. In contrast with already existing caching proxies, Squid for example, in a similar uploading scenario, the request would simply get redirected, leaving out potentially gains such us minimized processing power, lower server load and bandwidth. After prototyping and testing the suggested application against Squid, concerning data uploads with $1, 100, 1.000, \ldots, 100.000$ requests, we consistently observed query execution improvements ranging from 5 to 9 times. This enhancement was achieved through buffering and bulk-writing the data, the extent of which depended on the specific test conditions.}


\onecolumn \maketitle \normalsize \setcounter{footnote}{0} \vfill

\section{Introduction}
\label{sec:introduction}

The wide use of the internet by people around the world has posed scalability challenges for many businesses and service providers \cite{4}. Long response times or even inaccessibility is a factor that affects the revenues of web-centric companies, leading to lower earnings \cite{3}, \cite{4}. Web caches have been shown to solve some of the scalability problems. They helped bring down latencies, bandwidth usage and save processing power \cite{1}, \cite{3}, \cite{4}. 

There are generally a few widely used approaches to caching the data: browser cache, proxy cache and server cache \cite{1}, \cite{3}, \cite{5}, \cite{6}. The browser cache is the closest one to the user. It can save, in the memory of the local computer, static data like images, videos, CSS and JS code, etc. \cite{4}. A proxy cache is a dedicated server that sits between one or more clients and one or more servers. Compared to the browser cache, which is tied to a single machine, a proxy cache can be placed anywhere on the web, at different levels: ISP (local, regional, national) or right in front of the primary server \cite{4}. Lastly, there is the option to cache your data on the computer that is running the web server/database, either by using your own/a third-party solution or indirectly through the caching system of your operating system or database. 

Research in the field primarily targets cache replacement algorithms and prefetching. Crucially, caching solutions must decide what objects to retain and which to evict due to limited memory space. Managing the resources incorrectly and keeping unused objects cached for long enough, results in what’s known as cache pollution \cite{5}, \cite{9}. Some of the most popular caching policies include: LFU (least frequently used), LRU (least recently used), GDS (greedy dual size), GDS-Frequency and many more \cite{5}, \cite{6}, \cite{7}, some of them also using machine learning to enhance the already used ones \cite{12}. 

The effectiveness of caching is typically measured in hit rate or byte hit rate. Hit rates are determined as the percent of requests that could be satisfied directly by the cache, while byte hit rate represents the percent of the data (numbered in bytes) that were already cached before answering the request \cite{10}, \cite{11}. 

Cache prefetching is a technique that aims to request and cache objects before they are needed. Although useful, the solution must be carefully implemented to avoid polluting the cache in an excessive way \cite{5}, \cite{9}, \cite{22}. In their studies, \cite{5} showed that based on surveys, web caching along prefetching can reduce the latency of responses by up to 60\%, compared to web caching along, yielding an improvement of only 26\% in latency. Likewise, \cite{1} published a survey where we can find similar results, incorporating prefetching resulting in a 41\% or even 57\% latency improvement. 

The majority of the developed solutions for web caching are dedicated towards storing data generated by a server. Few of them offer a solution for buffering the incoming information that is meant to be stored by the SQL/NoSQL database. Indeed, Redis, Memcached, Apache Kafka, RabbitMQ and others offer the possibility to achieve this goal, but it is done intrusively, meaning that the underlying main server needs to suffer modifications to accommodate these solutions. In the following sections, we will introduce a system and architecture specifically engineered to cache incoming (uploaded) and fetched data, requiring minimal to no modifications to the core application. 

\section{Materials and Methods}
\label{sec:materials_and_methods}

Prior to delving into the construction and testing of the proposed solution, it is imperative to provide an overview of its counterpart. This would be Squid \cite{23}, a proxy server solution that is most often used as a caching proxy. It is a popular application used for caching and managing both static and dynamic content generated as response by a web server for a user’s request, supporting protocols such as HTTP, HTTPS, FTP and more. 

In order to save bandwidth, speed up load times, and conserve computing power, hundreds of Internet providers employ Squid in addition to thousands of standalone websites, as stated in \cite{23}. 

Squid is a battle-hardened application. It provides a powerful configuration file with the ability to create very complex distributed caching infrastructures. It can deploy on multiple computers and create a caching hierarchy consisting of Parents, Kids and Coordinators, all communicating with each other for better buffering and cache management to save bandwidth, processing power and lower latencies. Besides that, Squid also offers administrators the possibility to configure both Memory and Disk caches independently, each with its own replacement policy like: LRU, heap GDSF, heap LFUDA, heap LRU. 

Trying to match the power that Squid provides would be a very tedious and long process. Thus, we will try to optimize only a small chunk of it. Our focus falls on how Squid handles requests that upload data. As of now, Squid will simply redirect those to the main server. This approach may be improved. Instead of redirecting the request, we could extract the data (if its URL is marked as cacheable in the configuration file) and store it locally into a buffer till the caching time expires, sending it all at once afterward. This way, if the data is meant to be written in a relational database, the time it takes to execute for a single, multiple rows insertion query is far lower than overall multiple single rows insertions. The following sections will explore the concept further and present a viable solution. 

\begin{figure*}[!h]
  \centering
   {\epsfig{file = 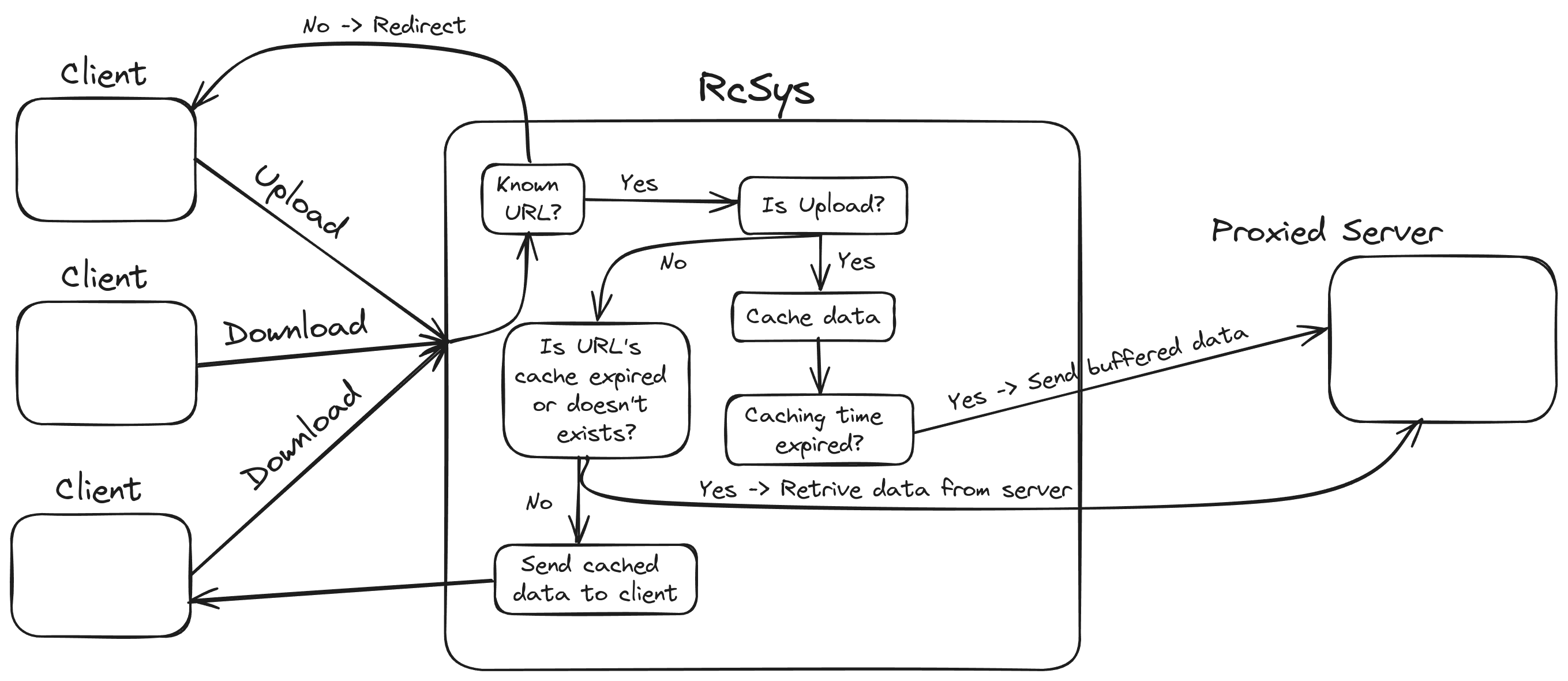, scale=0.5}}
  \caption{RcSys flow.}
  \label{fig:rcsys_flow}
 \end{figure*}
\section{The Proposed System Architecture}
\label{sec:the_proposed_system_architecture}
The proposed application, referred to as RcSys (Resource Caching System), functions as a caching proxy for handling both user upload and download requests. Prior to a detailed technical examination of the system's architecture, a general schematic overview of its operational flow will be presented. 

The diagram \ref{fig:rcsys_flow} illustrates three primary actors: clients, the RcSys server, and the main server. All communication between these components occurs through HTTP(S). 

The interaction commences as a client initiates a request over the internet, either for uploading (e.g., sending emails, creating posts) or downloading (e.g., reading messages, shopping). Upon reaching RcSys, a rapid evaluation occurs to determine if the requested resource should be cached. This decision relies on details outlined in a configuration file created by the administrator, specifying the paths designated for caching. 

If the requested URL is not known, then it gets redirected to the main server. In case it is known and is of type upload, it will be stored in a buffer and later, when the specified amount of caching time expires, it will be written to the main server (along with the other data that share the same URL). If the accessed resource is of type download, the system first checks whether it exists or is expired (if so, it makes a call for the updated version to the main server) and then replies to the client.

\subsection{Multi-Thread Request Processing}
Regarding the architecture of RcSys, a notable technical aspect involves request processing. It employs a dedicated Server thread for managing the web server and accepting connections, along with a pool of Worker threads controlled by a master thread. Upon a new connection, it becomes a task in a shared queue between Server and Worker. The master thread retrieves and assigns tasks to workers. Simultaneously, new connections can be established and added to the queue. 

\begin{figure*}[!h]
  \centering
   {\epsfig{file = 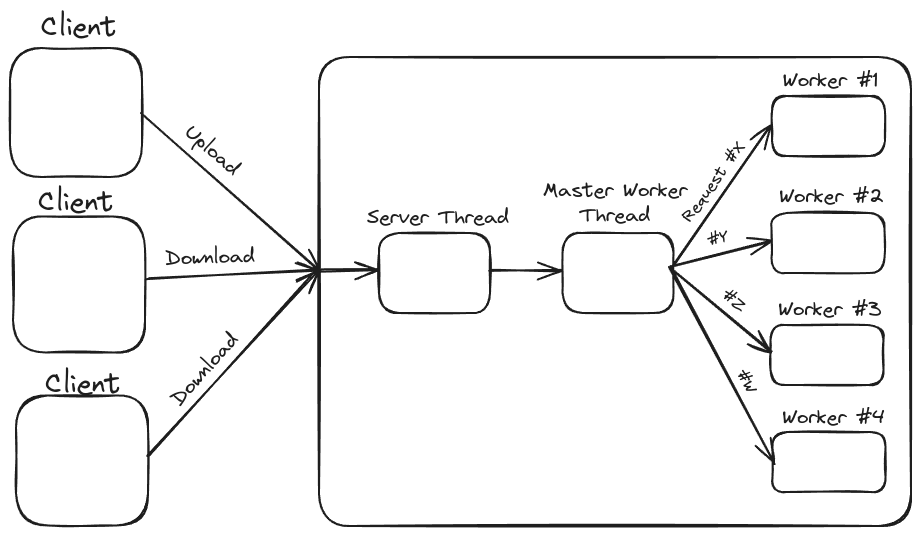, scale=1}}
  \caption{RcSys thread pool.}
  \label{fig:rcsys_thread_pool}
 \end{figure*}

The Figure \ref{fig:rcsys_thread_pool} offers a more detailed view of the system's functionality. Upon the application's initiation, a configuration file undergoes processing, and its supplied information is stored within an IConfiguration object for convenient access. This file encompasses various details, including the desired thread pool size. Concurrently, the creation of the Worker results in the instantiation of X threads, as dictated by the configuration. 

This architectural decision is predicated on a challenge encountered during the design phase of the proxy caching system. To illustrate, consider a scenario where a singular Worker thread manages user requests. In a context where numerous resources are cached, each with its specific expiration time ranging from seconds to hours, and a substantial number of these resources total in the hundreds. For instance, if there are 1000 requests awaiting processing, some requiring the update of cached data, the consequence is that over 900 requests must wait for the arrival of new information, irrespective of whether it pertains to their specific data or not. In contrast, utilizing a thread pool, the proposed architecture enables concurrent processing of new requests by the CPU while the operating system monitors the arrival of necessary resources from the main server to fulfill other pending tasks. 

The requests are distributed for processing among the worker threads in a Round-robin fashion \cite{14}. An index of the thread that is next for receiving a new task is kept in the Master Worker state. If a new request arrives, the thread that is pointed at by the index will be assigned to serve it and the index is incremented, therefor the next one will be handled by a different worker. 

A second solution for solving the earlier mentioned problem would be to build an event-loop based architecture. However, such a resolve imposes new challenges and a harder to comprehend implementation, even if it is used a library like libev\cite{24}. The more natural way is to use threads, performing as good as event-loop programming and easier to execute \cite{13}. 

\subsection{One-Tiered Cache Replacement Policy}
The main mean by which RcSys stores the cached data is disk, be it HDD or SSD. Although it may not be an option as faster to access and retrieve information as RAM, it has its own advantages and the fact that it is “disk-only” is not entirely true. 

Caching data into RAM comes with a lot of careful managing and designing. If your process ends up leaking memory or not allocating/deallocating it efficiently, it can use all computer’s memory and slow the entire system performance or even crash. RAM is also a much smaller sized resource on commodity computers that can be used as servers, being significantly more expensive than the disk. A $16-32-64-128 \ldots GB$ of RAM machine could also store way less cached data compared to a 2-4TB non-volatile memory option. 

Despite its inherent drawbacks, the decision to utilize the disk as a storage medium is justified by the support provided by the operating system in managing file access. Therefore, the characterization of it being a "disk-only solution" is not entirely accurate. In order to facilitate access to memory for software applications, the OS allocates pages of memory. Those pages represent virtual memory addresses that are later translated to physical addresses when a request to the memory controller is made to get the stored data. The OS also uses the main memory of a computer to load accessed disk files into and minimize the IO operations that would be performed. It maps the file opened for reading and writing from the disk to the RAM and takes care of evicting them if the memory is full. This way, the user can manipulate the file’s data and the system does not have to issue a write to disk every time a new letter is typed. Instead, it marks the memory pages as “dirty” and updates the file later. Managing resources this way increases the machines’ general performance \cite{15}. 

RcSys’s architecture takes advantage of the OS file caching behavior to simplify the proposed solution and to also benefit from the performance gains that come along writing and reading from RAM. A similar approach is implemented by Nginx for caching, as detailed by one of their Senior Director here \cite{16}. 

Disk’s higher memory availability does not mean that it is infinite. It might as well get full if enough data requires caching. To deal with this problem, a simple cache eviction policy was implemented, that would delete files from disk according to LRU. The least recently accessed file for either reading or writing will be erased to make space for a new one. The configuration file also specifies the maximum size of disk memory that can be used for caching. The architecture, however, does not guarantee that the quantum won’t be exceeded. Instead, when a new request comes in, before processing it, we evaluate whether the maximum cache size was exceeded. If so, cached resources from the disk will get deleted in a LRU manner until the used space is below the maximum one. This means that if the incoming request has to write new files on disk, the memory limit could be again surpassed, till another request arrives and the cycle repeats. The OS also implements a variation of the LRU for deleting unused pages \cite{17}. 

\subsection{Use Cases}
As a caching proxy, RcSys aligns with established use cases observed in existing solutions. Its deployment serves purposes such as conserving processing power and bandwidth for a web server. By caching non-real-time critical content, RcSys alleviates the proxied server's burden, including items like blog posts, images, messages, entire web pages, etc. Moreover, it enhances the user experience by positioning itself in proximity to the user, handling requests that do not necessitate database queries or data regeneration, thereby minimizing overhead. 

An additional advantage that sets it apart from alternative solutions is its capability to temporarily store data intended for server writes and subsequently perform bulk writes. This feature substantially reduces the processing load on the main server and database, further enhancing efficiency. 

\subsection{A Final Overview of System's Flow}
\begin{figure*}[!h]
  \centering
   {\epsfig{file = 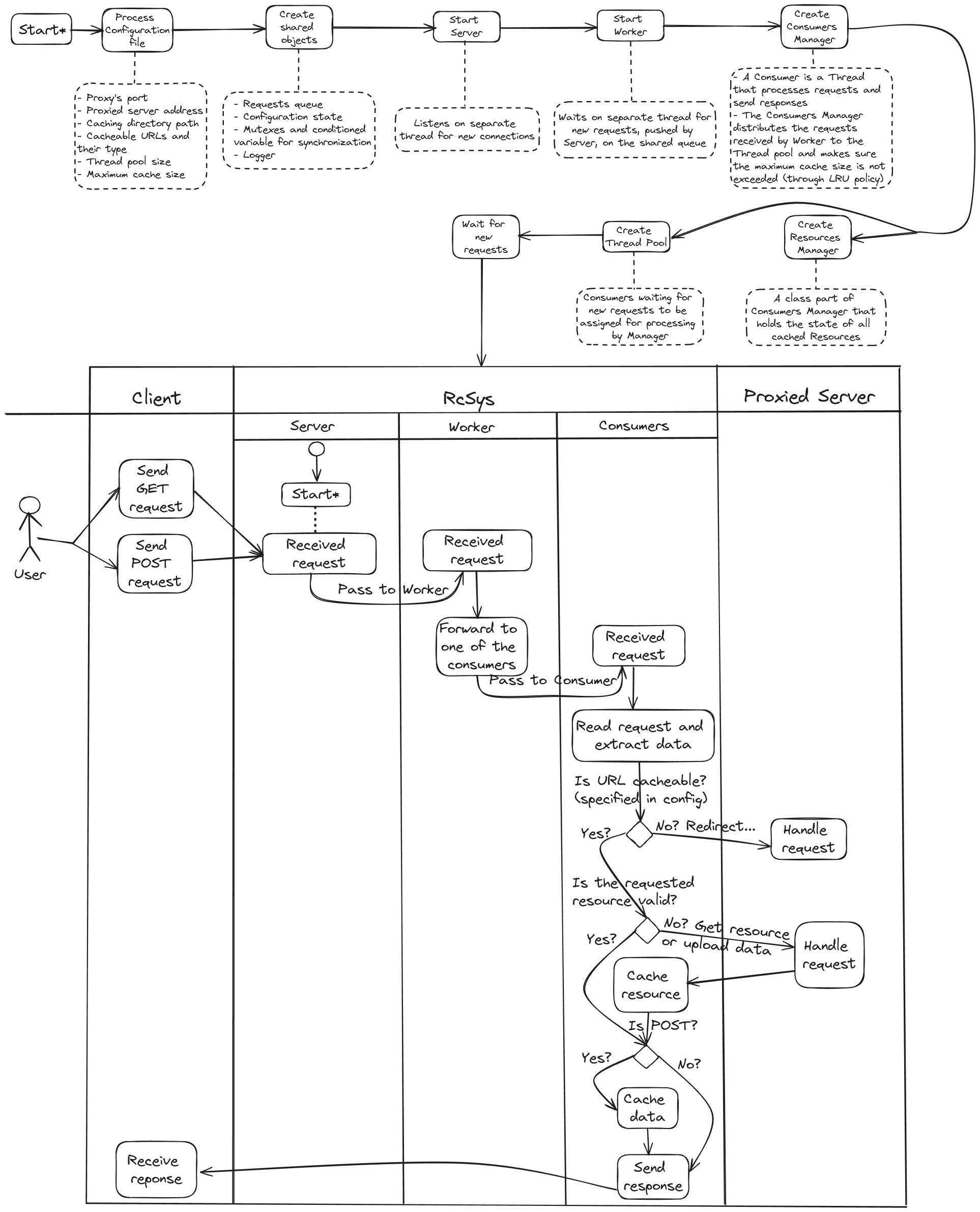, scale=0.8}}
  \caption{RcSys flow chart.}
  \label{fig:rcsys_flow_chart}
 \end{figure*}

For providing a comprehensive overview and connecting all the components comprising the architecture, a sequence diagram has been crafted, see Figure \ref{fig:rcsys_flow_chart}. In this diagram, we explain some of the most important objects' states and their role in RcSys, as well as showing the path that a user’s request takes once in the system.

\begin{figure*}[!h]
  \centering
   {\epsfig{file = 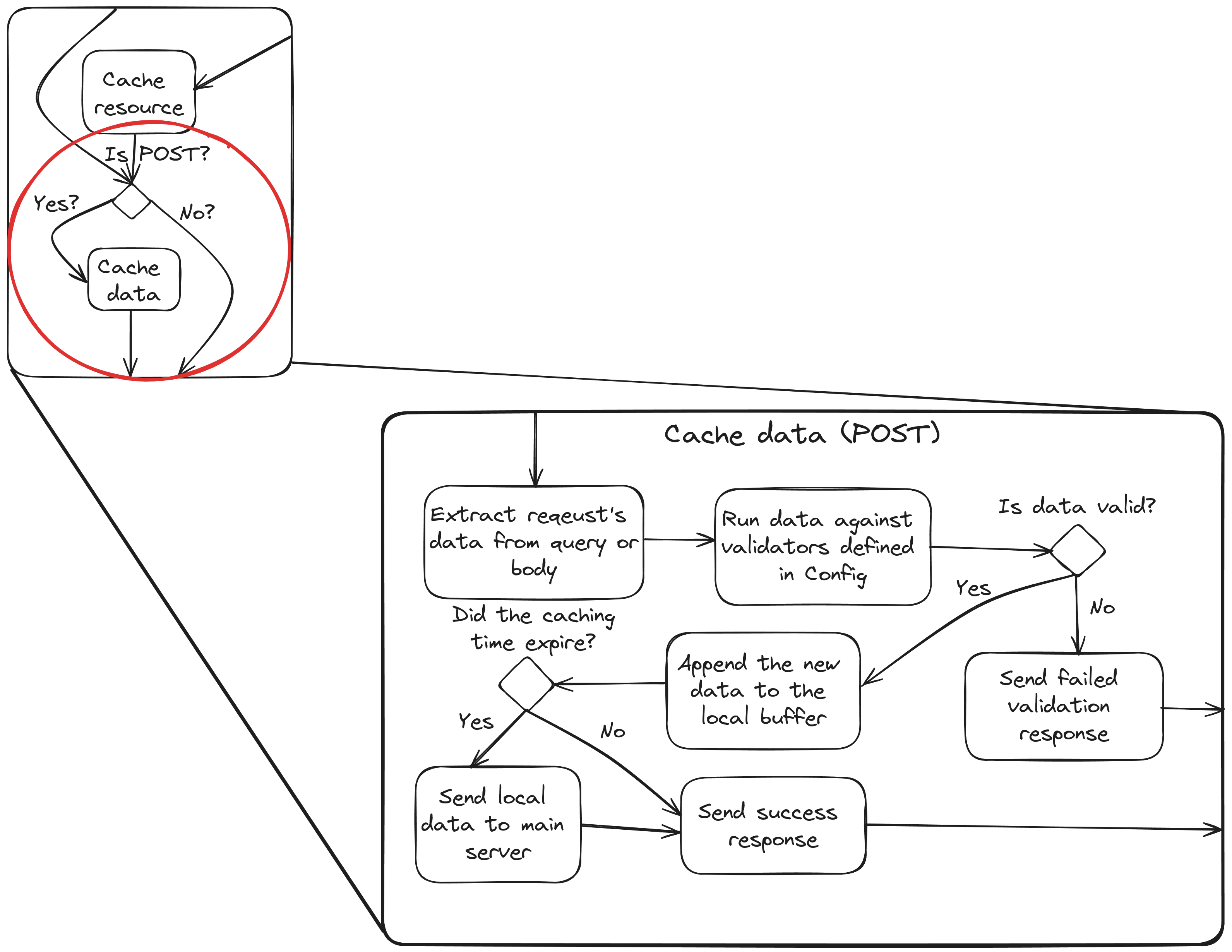, scale=0.35}}
  \caption{RcSys flow chart zoomed in on POST cache.}
  \label{fig:rcsys_zoom_in_cache}
 \end{figure*}

If we delve into the details, particularly in the context of caching incoming data, the process unfolds as follows in \ref{fig:rcsys_zoom_in_cache}. 
\begin{figure*}[!h]
  \centering
   {\epsfig{file = 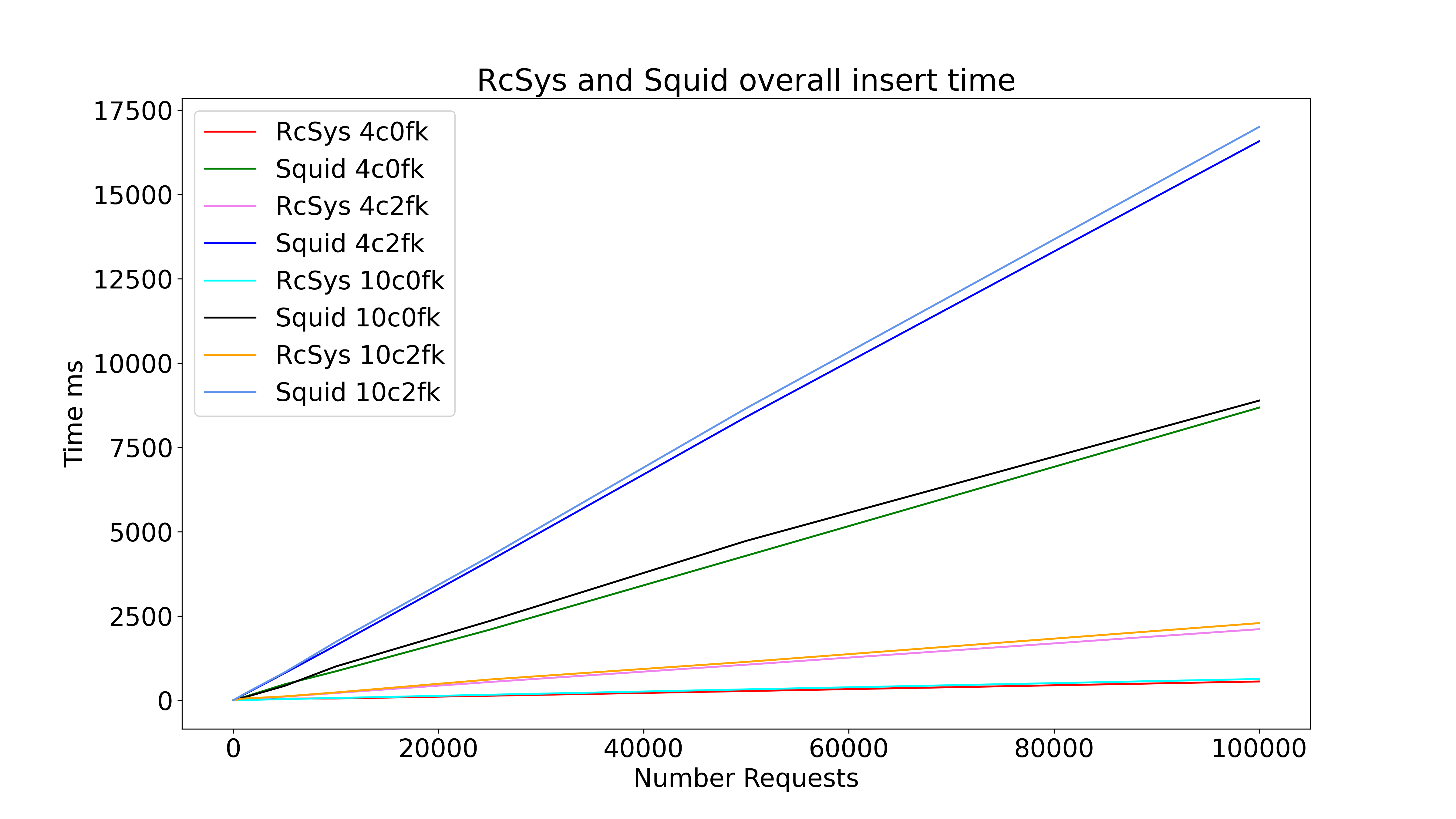, scale=0.25}}
  \caption{Overall performance gains RcSys vs Squid.}
  \label{fig:overall_gains}
 \end{figure*}

\section{Strengths and Shortcomings of the Solution}
In order to compare the overall performance of the proposed application, we chose to test it against Squid-Cache \cite{23}, both being a caching proxy. Before going forward, we must make a big disclaimer. RcSys is nowhere near as powerful and well-rounded as squid, nor is it production ready. It is just an experimental application built to explore new architectures and capabilities that a caching proxy can bring. 

The proxied application utilized in our research serves as a conceptual, abstract representation—a pattern mirroring the structure typical of a conventional web server. Developed in Golang, the choice of this language stems from our desire for a swift prototype creation process, minimizing unnecessary code verbosity. 

Golang’s net/http \cite{25} handles the requests along Gorm \cite{26} for interacting with a PostgreSQL database. The database’s schema is modest, consisting of four entities (4c0fk, 4c2fk, 10c0fk, 10c2fk), modeled in such a way as to cover some real-world operations. The names are self-explanatory, each entity consisting of several string fields equal to the digits previous to the "c” character and some foreign keys equal to the digit previous to “fk”. For example, entity 4c0fk has 4 string columns and 0 foreign keys, while entity 10c2fk has 10 string columns and 2 foreign keys (mapped to entity 4c0fk and entity 10c0fk, same as for entity 4c2fk). 

We aimed to explore fundamental database use cases by conducting tests for different results on tables of various sizes, ranging from small to large, and considering scenarios with and without foreign keys. Foreign keys play an important role in a relational database, impacting a query's performance. For each foreign key that the database must insert, it has to perform a look up in the referenced table to make sure that the new created row is valid and does not point to a non-existing record. 

The application that simulates clients sending requests to the web server was also built with Golang for the same fast-prototyping reason. It defines basic functions for getting as well as posting data for several times in order to create metrics and compare the proxies.  

For the first set of tests, we simulated basic POST operations, with the requests being proxied by both RcSys and Squid. As already mentioned, Squid does not possess the ability to cache the uploaded data in order to send it later to the proxied server all at once. Therefore, every such operation was redirected to the main server. Both applications went through the same tests, with the same data (different strings with the same size). For each entity (4c0fk, 4c2fk, 10c0fk, 10c2fk), a client sent $1, 100, 1000, 5000, 10000, 25000, 50000$ and 100000 POST HTTP requests to create a new record in the database. The data was collected and analyzed for each batch of tests $(1, 100, \ldots)$. In order to obtain accurate results of the execution time, PostgreSQL provides us with a query prefix EXPLAIN ANALYZE \cite{18}, \cite{19}, \cite{20}. Running a simple query like "INSERT INTO ... VALUES ..."
and prefixing it with "EXPLAIN ANALYZE =\textgreater EXPLAIN ANALYZE INSERT INTO ... VALUES ... "
results in both executing the query and outputting different real information about the execution process. From among the returned data, we extracted two performance metrics, namely the "Execution time" (the actual time that it took to execute the query) and "Planning time" (the actual time that it took to plan the query execution), expressed in milliseconds. The information that EXPLAIN ANALYZE provides differs for each type of query. For example, if the inserted data links through foreign keys to other tables, a trigger will be executed to check if the referenced row exists. This "Trigger time" is summed up in the "Execution time" parameter. 

Figures \ref{fig:overall_gains} compare the execution and planning time data for all mentioned entities, the $X$ axis being the number of requests $1, 100, 1.000, 5.000, \ldots 100.000$ and the $Y$ axis being the execution time corresponding to the $X$ number of requests. Observing these results, it becomes evident that RcSys outperforms Squid significantly in handling upload requests. Saving multiple rows of data is faster than saving only one at a time (one per query, as Squid did) and significantly mitigates the overall inserting cost. Although RcSys could buffer all the incoming upload requests in a span as long as the caching expiration time, we chose to limit the amount of data to as much as 10.000 requests, after which the buffer would be emptied. MySQL’s documentation \cite{21} breaks down the cost of an INSERT statement in proportions as follows: 
\begin{itemize}
    \item Connecting (3)
    \item Sending query to server (2)
    \item Parsing query (2)
    \item Inserting row (1 x size row)
    \item Inserting indexes (1 x number of indexes) 
    \item Closing (1)
\end{itemize}. 
Inserting multiple rows at once allowed avoiding the costs associated with multiple connections, sending queries to the server, and closures. Examination of the underlying data used in generating the four charts reveals the corresponding tables, see Tables \ref{tab:raw_tests_data_execution} and \ref{tab:raw_tests_data_planning}.
\begin{table*}[!h]
\tiny
\caption{Raw execution tests data RcSys vs Squid in ms.}\label{tab:raw_tests_data_execution} \centering
\begin{tabular}{|c|c|c|c|c|c|c|c|}
  \hline
    \multicolumn{2}{|c|}{Entity} & 1 & 100 & 1.000 & 10.000 & 50.000 & 100.000 \\
  \hline 
  \multirow{2}{*}{4c0fk}
     & RcSys & 0.055 & 1.831 & 9.64 & 48.106 & 254.99 & 514.69  \\
     & Squid & 0.055 & 8.249 & 65.345 & 633.505 & 3174.3 & 6413.4 \\
  \hline 
  \multirow{2}{*}{4c2fk}
     & RcSys & 0.313 & 2.348 & 20.379 & 198.445 & 984.092 & 1956.51  \\
     & Squid & 0.313 & 11.431 & 131.999 & 1314.13 & 6791.7 & 13395.5 \\
  \hline 
  \multirow{2}{*}{10c0fk}
     & RcSys & 0.299 & 0.758 & 6.368 & 59.859 & 296.108 & 578.355  \\
     & Squid & 0.299 & 6.11 & 57.113 & 730.357 & 3446.2 & 6458.6 \\
  \hline 
  \multirow{2}{*}{10c2fk}
     & RcSys & 0.281 & 2.781 & 45.473 & 211.247 & 1058.82 & 2121.50  \\
     & Squid & 0.281 & 13.23 & 142.361 & 1396.49 & 6979.6 & 13692.9 \\
  \hline
\end{tabular}
\end{table*}
\begin{table*}[!h]
\tiny
\caption{Raw planning tests data RcSys vs Squid in ms.}\label{tab:raw_tests_data_planning} \centering
\begin{tabular}{|c|c|c|c|c|c|c|c|}
  \hline
    \multicolumn{2}{|c|}{Entity} & 1 & 100 & 1.000 & 10.000 & 50.000 & 100.000 \\
  \hline 
  \multirow{2}{*}{4c0fk}
     & RcSys & 0.019 & 0.101 & 0.318 & 4.621 & 17.75 & 41.99  \\
     & Squid & 0.019 & 2.988 & 25.141 & 225.035 & 1112.1 & 2265.0 \\
  \hline 
  \multirow{2}{*}{4c2fk}
     & RcSys & 0.024 & 0.124 & 1.095 & 17.235 & 70.487 & 150.85  \\
     & Squid & 0.024 & 2.639 & 30.298 & 311.137 & 1611.1 & 3182.5 \\
  \hline 
  \multirow{2}{*}{10c0fk}
     & RcSys & 0.026 & 0.052 & 0.32 & 7.581 & 26.919 & 52.237  \\
     & Squid & 0.026 & 2.365 & 21.413 & 272.29 & 1280.3 & 2427.8 \\
  \hline 
  \multirow{2}{*}{10c2fk}
     & RcSys & 0.022 & 0.464 & 4.161 & 20.39 & 79.134 & 165.84  \\
     & Squid & 0.022 & 3.134 & 33.755 & 335.182 & 1679.7 & 3310.1 \\
  \hline
\end{tabular}
\end{table*}

Extracting additional information about the gains is possible from the raw data, such as the average time improvement for each test, as indicated in Table \ref{tab:performance_gains}. Besides its advantage in caching upload requests, RcSys lacks many features and critical functionalities that Squid has, see Table \ref{tab:rcsys_vs_squid}.

\begin{table*}[!h]
\tiny
\caption{RcSys vs Squid performance gains.}\label{tab:performance_gains} \centering
\begin{tabular}{|c|c|c|c|c|}
  \hline
  Average query speedups & Test 4c0fk & Test 10c0fk & Test 4c2fk & Test 10c2fk \\
  \hline
  Execution Speedup & x9.12 & x9.48 & x5.93 & x5.20 \\
  \hline
  Planning Speedup & x52.11 & x40.18 & x19.13 & x14.17 \\
  \hline
\end{tabular}
\end{table*}

\begin{table*}[!h]
\tiny
\caption{RcSys vs Squid features comparison.}\label{tab:rcsys_vs_squid} \centering
\begin{tabular}{|c|c|c|}
  \hline
  Feature & RcSys & Squid \\
  \hline
  Security & No HTTPS (SSL/TLS) or Authentication & HTTPS and Authentication \\
  \hline
  Protocols supported & Only HTTP & HTTP, HTTPS, FTP and more \\
  \hline
  Distributed cache & No & \shortstack{Yes, with complex hierarchy of \\ parents, kids and coordinators} \\
  \hline
  Cache replacement policies & LRU & \shortstack{LRU, heap GDSF, \\ heap LFUDA, heap LRU} \\
  \hline
  Caching options & Only disk & Disk and Memory \\
  \hline
  Configuration & Minimal & Very robust \\
  \hline
\end{tabular}
\end{table*}

\section{Conclusions}
The paper has proposed an innovative enhancement designed to bring added value to conventional caching proxy solutions. Through the implementation and testing of the proposed architecture, it was evident that there is a significant improvement opportunity in reducing bandwidth consumption and optimizing data upload speeds, especially in contexts utilizing relational databases. The strategy of locally extracting and storing data from upload-type requests on the caching server's disk or memory, followed by later bulk-write, showcased marked enhancements in overall database write performance. While typical web caching proxies would redirect those types of requests, RcSys obtained by buffering them between 5.20 and 9.12 times SQL query execution speedups and between 14.17 and 52.11 times SQL query planning speedups. 

\bibliographystyle{apalike}
{\small
\bibliography{example}}

\end{document}